\documentclass[prl,twocolumn,superscriptaddress,nopacs,preprintnumbers,amsmath,amssymb,floatfix]{revtex4}
\usepackage[colorlinks=true, linkcolor=blue, citecolor=blue]{hyperref}
\usepackage{epsfig}

\usepackage{color}
\begin{document}

\title{CDW slips and giant frictional dissipation peaks at the NbSe$_2$ surface}

\author{M. Langer}
\affiliation{Department of Physics, University of Basel,Klingelbergstr. 82, 4056 Basel, Switzerland}
\author{M. Kisiel}
\affiliation{Department of Physics, University of Basel,Klingelbergstr. 82, 4056 Basel, Switzerland}
\author{R. Pawlak}
\affiliation{Department of Physics, University of Basel,Klingelbergstr. 82, 4056 Basel, Switzerland}
\author{F. Pellegrini}
\affiliation{SISSA, Via Bonomea 265, 34136 Trieste, Italy}
\affiliation{CNR-IOM Democritos National Simulation Center, Via Bonomea 265, I-34136 Trieste, Italy}
\author{G.E. Santoro}
\affiliation{SISSA, Via Bonomea 265, 34136 Trieste, Italy}
\affiliation{CNR-IOM Democritos National Simulation Center, Via Bonomea 265, I-34136 Trieste, Italy}
\affiliation{International Centre for Theoretical Physics (ICTP), P.O.Box 586, I-34014 Trieste, Italy}
\author{R. Buzio}
\affiliation{CNR-SPIN Institute for Superconductivity, Innovative Materials and Devices, C.so Perrone 24, 16152 Genova, Italy}
\author{A. Gerbi}
\affiliation{CNR-SPIN Institute for Superconductivity, Innovative Materials and Devices, C.so Perrone 24, 16152 Genova, Italy}
\author{G. Balakrishnan}
\affiliation{Department of Physics, University of Warwick, Coventry CV4 7AL, UK}
\author{A. Baratoff}
\affiliation{Department of Physics, University of Basel,Klingelbergstr. 82, 4056 Basel, Switzerland}
\author{E. Tosatti}
\affiliation{SISSA, Via Bonomea 265, 34136 Trieste, Italy}
\affiliation{CNR-IOM Democritos National Simulation Center, Via Bonomea 265, I-34136 Trieste, Italy}
\affiliation{International Centre for Theoretical Physics (ICTP), P.O.Box 586, I-34014 Trieste, Italy}
\author{E. Meyer}
\affiliation{Department of Physics, University of Basel,Klingelbergstr. 82, 4056 Basel, Switzerland}

\date{\today}

\begin{abstract}
Accessing, controlling and understanding nanoscale friction and dissipation is a crucial issue 
in nanotechnology, where moving elements are central
\cite{Urbakh2010, Lantz2009, Socoliuc2006, Vanossi2013}.  
Recently, ultra-sensitive noncontact pendulum Atomic Force Microscope (AFM) succeeded 
in detecting the electronic friction drop caused by the
onset of superconductivity in Nb \cite{Kisiel2011}, raising hopes that a wider variety of mechanisms 
of mechanical dissipation arising from electron organization into different collective 
phenomena will become accessible through this unconventional
surface probe. Among them, the driven phase dynamics 
of charge-density-waves (CDWs) represents an outstanding challenge as a source of dissipation. 
Here we report a striking multiplet of AFM dissipation peaks arising at nanometer distances above the 
surface of NbSe$_2$ - a layered compound exhibiting an incommensurate CDW. 
Each peak appears at a well defined tip-surface interaction force of the order of a nN, 
and persists until T~=~70~K where CDW short-range order is known to disappear.  
A theoretical model is presented showing that the  peaks are connected to tip-induced local 2$\pi$ 
CDW phase slips. Under the attractive potential of the approaching tip, the local CDW surface phase landscape deforms 
continuously until a series of 2$\pi$ jumps occur between different values of the local phase.
As the tip oscillates to and fro, each slip gives rise to a hysteresis cycle, appearing at a selected distance, the dissipation  
corresponding to ``pumping''  in  and out a local slip in the surface CDW phase of  NbSe$_2$.
\end{abstract}

\maketitle

Bodies in relative motion separated by large vacuum gaps of one or more nanometers 
experience a tiny frictional force, whose nature is now beginning to be
accessible \cite{Volo2007}.  Measuring these minute frictional forces, and relating them to the
underlying physics and collective phenomena of electrons, ions, spins and their phase transitions
represents a challenging and valuable prospective form of local spectroscopy of the solid surface under the tip. 
Although extremely delicate, this noncontact form of friction 
is now measurable by a highly sensitive AFM cantilever oscillating like a
pendulum over the surface \cite{Stipe2001, Kuehn2006, Kisiel2011}. The pendulum configuration takes advantage of very small spring
constants, typically in the order of mN/m. The corresponding minimal detectable
noncontact friction is extremely small, about $\Gamma_0^{\rm pend} \sim 10^{-12}$~kg/s,
compared with $\Gamma_0^{\rm TF} \sim 10^{-7}$~kg/s of more conventional tuning fork tips
(see Methods for details),  which work at smaller distances, just outside the repulsive regime.
Here we use the pendulum to demonstrate  a novel example of noncontact friction,
occurring when the vibrating tip pumps integer 2$\pi$ slips onto the local surface phase of a 
CDW. \\

Layered dichalchogenides have long been known for their phase transitions leading to
picometer sized superstructure lattice distortions and corresponding new electronic periodicities 
in their low temperature ground state\cite{Wilson1975}.   
Among them, NbSe$_{2}$ (with 2H stacking) stands out 
as a material exhibiting bulk CDW below the long range order onset temperature 
$T_{\rm CDW}=33$~K \cite{Moncton1975}, 
as well as superconductivity below $T=7.2$~K. The CDW consists of a joint periodic lattice distortion (PLD) 
and electron density modulation, with the same incommensurate periodicity close to $(3\times3)$ 
in the layer plane, and a remarkably coherent phase, extending over hundreds of lattice spacings 
in clean samples. 
Whereas in other isoelectronic 
compounds of that family, such as TaS$_2$ (with 1T stacking), the CDW has a clear
connection with the Fermi surface, thus leading to an insulating or near insulating state
at low temperatures~\cite{Wilson1975}, the driving force in NbSe$_2$ 
is now known to be different, not directly involving the Fermi surface, as recently ascertained both
experimentally \cite{Inosov2008, Shen2008, Borisenko2009, Weber2011, Pasupathy2013}  
and theoretically \cite{Weber2011}. 
The distortion of NbSe$_2$ is thus really only a PLD. Nonetheless
we stick to the commonly used term ``CDW'' to avoid language confusion. 

\begin{figure}[htb]
\centering
\includegraphics[width=0.48\textwidth]{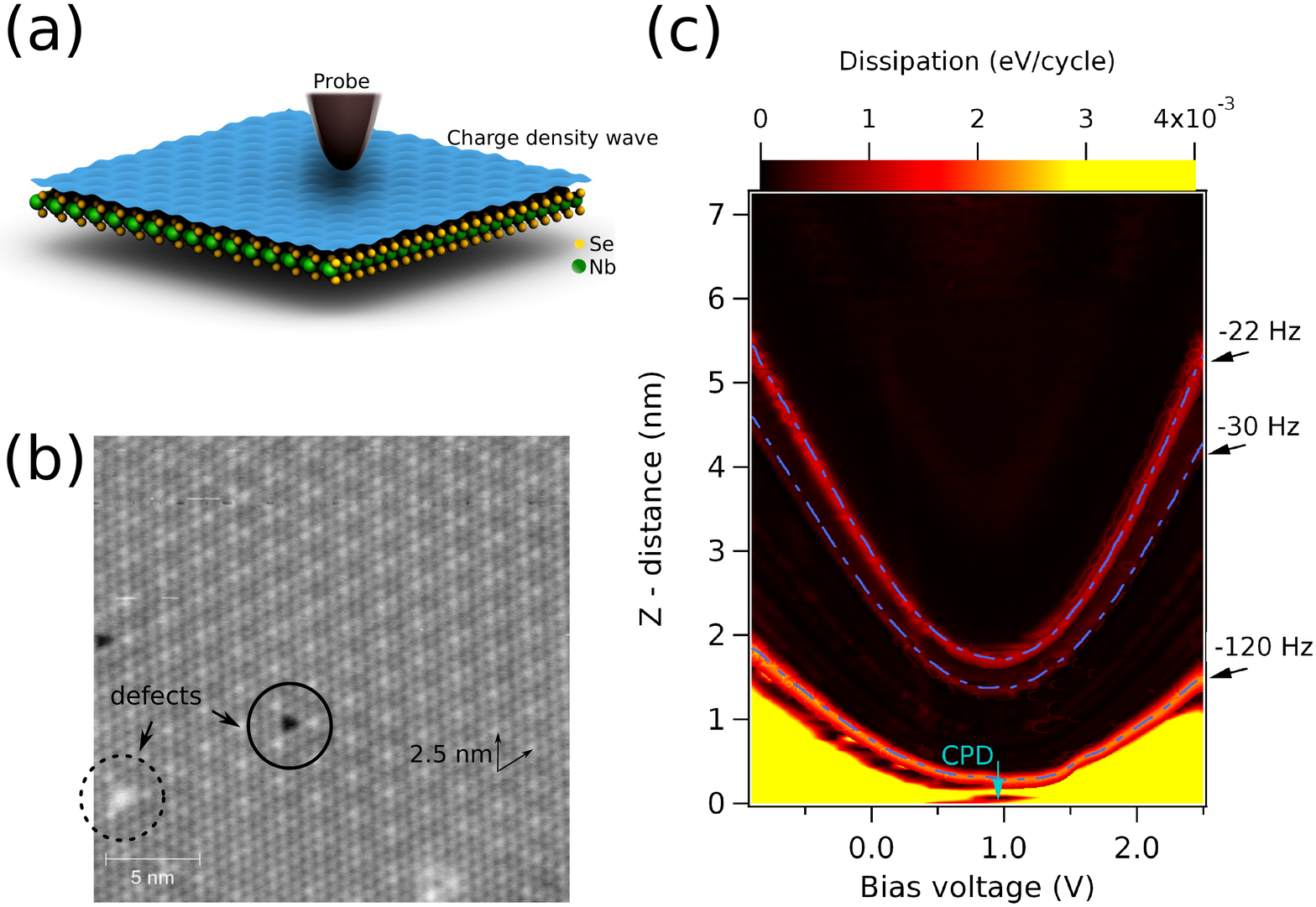}
   \caption{Observation of Charge density wave on NbSe$_2$ surface
and accompanied noncontact friction. \textbf{a}, an oscillating AFM tip in
proximity to the Charge Density Wave on NbSe$_2$ surface. \textbf{b}, constant
current STM ($I=10$~pA, $V=5$~mV) image of NbSe$_2$ surface showing hexagonal CDW
induced Moir\'e pattern as well as two types of surface defects - adsorbed CO
molecules (dashed circle) and Se atom vacancies (circle). \textbf{c}, energy
dissipation between NbSe$_2$ surface and the pendulum AFM tip \textit{versus}
tip - sample distance ($Z$) and bias voltage $V$. Bright features on the image stand
for large noncontact friction. Friction increase is at the same cantilever
frequency $f=f_0-22$~Hz, $f=f_0-30$~Hz, $f=f_0-120$~Hz, and the constant
frequency contours are shown with dashed lines. 
The distance d=0 corresponds to the point where the tip enters the contact regime, which we clearly see in cantilever deflection signal. 
Measurement was acquired at $T=6$~K.}
   \label{fig1}
\end{figure}

We measured the friction force acting between an \textit{in situ} cleaved
NbSe$_{2}$ surface and a sharp silicon cantilever tip oscillating in the
pendulum geometry as well as a qPlus tuning fork tip. We simultaneously acquired the
tip oscillation frequency $f$  and a dissipation signal $P$ under
ultrahigh-vacuum (UHV) conditions and low temperatures of $T=5-6$~K, 
at which CDW and superconductivity coexist.
As a function of tip-surface distance, we observed a remarkable train of
dissipation maxima which extend as far out as a few nm above the surface. 
In order to understand their origin we consider a theoretical model, where the CDW
is treated as an elastic medium coupled to the spatially extended oscillating
cantilever tip (Fig.~\ref{fig1}(a)). A comparison of AFM data with theory allows us to
conclude that the series of dissipation maxima is due to a hysteretic behavior
of the CDW phase as the tip oscillates at specific distances where sharp local slips occur. 

We first obtained, by means of a standard tuning fork STM, the atomically resolved surface
topography showing the additional CDW induced Moir\'e pattern as shown
in Fig.~\ref{fig1}(b). The CDW modulation is incommensurate with the underlying
lattice, as is well known \cite{Moncton1975, Guillamon2008}. Depending on the chosen spot, 
the CDW modulation is known to exhibit its incommensurate deviation from the nearby $(3\times3)$ commensurate periodicity
either along one, or simultaneously along all three, planar directions \cite{Soumyan2013}. 
In the proximity of the few visible surface defects (CO adsorbates and Se vacancies) the CDW distortion, 
pinned by the impurity, survives well above the pristine $T_{\rm CDW}$. 

In Fig.~\ref{fig1}(c) the noncontact friction $P(z,V)$ \textit{versus}
tip-sample distance $z$ and tip-sample voltage $V$ as the tip approaches the
sample surface is shown for the pendulum AFM. The dissipated power in
units of electron volts per cycle is calculated according to a standard formula,
measuring the difference between the power in (first term)  and out (second term) of the tip\cite{Cleveland1998}:
\begin{equation}\label{DissPow}
P =  P_0 \left( \frac{A_{\rm exc}(d)}{A_{\rm exc,0}} - \frac{f(d)}{f_0}\right) \;,
\end{equation}
where $P_0$ is the dissipated power due to intrinsic 
losses of the cantilever, measured at large tip-sample separation, 
$A_{\rm exc}(d)$ and $f(d)$ are the distance dependent excitation amplitude 
(as measured by the excitation voltage needed to excite the cantilever at constant oscillation amplitude $A$) 
and frequency of the cantilever and the suffix zero refers to the free cantilever. In the case of
pendulum AFM, the frequency and the amplitude of lateral oscillations are equal
to $f_0=12$~kHz and $A=5$~nm, respectively.
Owing to the tip asymmetry (see Methods section and Supplementary material), there is a non negligible
perpendicular oscillation amplitude $A_{\rm perp} = 180$~pm as well. Bright
features on Fig.~\ref{fig1}(c) correspond to high dissipation maxima up to
$P=2$~meV/cycle observed already at some nm above the sample surface. 
Within each dissipation branch the amount of energy loss stays constant, independently of the
bias voltage $V$. The giant noncontact friction maxima remain well defined even after
careful compensation of Contact Potential Difference (CPD) between tip and
sample. The peaks persist whether the tip-surface interaction force $F$ is 
van der Waals ($V=V_{\rm CPD}=+0.95$~V) or electrostatic ($V\neq V_{\rm CPD}$). Moreover, 
the $z(V)$ voltage dependence of each dissipation branch has a parabolic behavior (as expected for
an electrostatic capacitive interaction between a roughly conical tip and sample, where $F \sim V^2/d$) implying that
each dissipation maximum always occurs at the same tip-surface interaction force as the voltage changes. 
Thus the effect is force controlled rather than voltage controlled.

\begin{figure}[htb]
\centering
\includegraphics[width=0.48\textwidth]{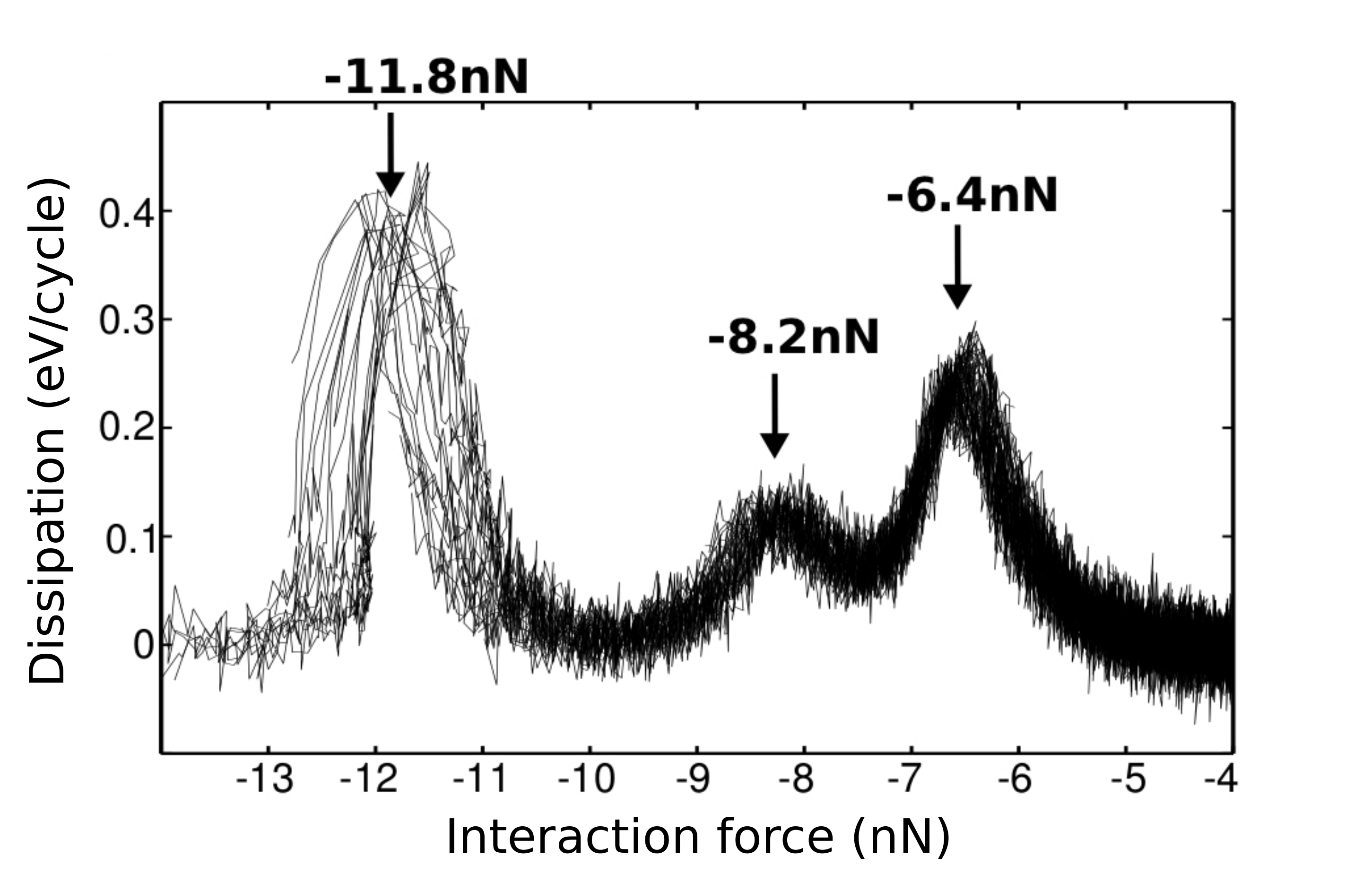}
   \caption{The energy dissipation \textit{versus} tip-sample
interaction force $F_{\rm int}$. Several spectra for different bias voltage $V$
applied between tuning fork tip and NbSe$_2$ surface are superimposed. Three
dissipation maxima appear always at the same $F_{\rm int}$, independent on $V$. 
Measurement was acquired at $T=5$~K.}
   \label{fig2}
\end{figure}

So far the pendulum AFM results. 
To further confirm and ascertain the universal nature of this new effect, we repeat
the experiment on an alternative instrument, the tuning fork AFM. In this setup, the tip oscillation 
direction is perpendicular to the sample surface, making the measurement more localized and 
ultimately more consistent across the surface. Fig.~\ref{fig2} shows the 
non contact friction of this different probe,
again as a function of $V$ ($-0.4$~V~$<V<1.5$~V), and the dissipated power again calculated 
according to equation (\ref{DissPow}), where $f_0=25$~kHz and $A=200$~pm. Tuning fork results
systematically show three reproducible dissipation maxima positioned at different interaction forces
$F_{\rm int}=-6.4$~nN, $-8.2$~nN, $-11.8$~nN as the tip approaches the surface (distance $0$~nm~$<d<2$~nm and the attractive force was 
extracted from the
cantilever frequency data by means of the Sader-Jarvis algorithm \cite{sader2004}). 
The finding of sharp dissipation peaks, a single one of which was already reported 
by Saitoh \cite{Saitoh2010}, is contrary to most noncontact
friction experiments, where the energy dissipation $P$ increases
smoothly as the interaction force rises \cite{Kisiel2011, Volo2007, Stipe2001,
Kuehn2006}. In particular, tip dissipation on a NbS$_2$ sample, yet another layered dichalcogenide compound, 
but without CDW, shows only a smooth increase (see Supplementary material). In summary
two independent UHV experiments - the pendulum AFM (Fig.~\ref{fig1}(c)) and -- with much smaller intensity-- 
the tuning fork AFM (Fig.~\ref{fig2}) - find a sequence of dissipation peaks 
extending as far out as several nm above the surface. 

What is their origin? Even before any theory, temperature provides a suggestion that the dissipation peaks are related to
the coupling of the tip to the CDW distortion. 
The dissipation peaks in fact survive up to a temperature of about 
$T=71\pm6$~K (see Fig.~3 of the Supplementary Materials). As mentioned earlier, STM data show 
that short-range CDW order in NbSe$_2$ persists around surface defects or
impurities at temperature up to $T=(2.5-3)T_{\rm CDW}$ \cite{Pasupathy2013}. 
Up to that temperature, the tip potential itself may thus awaken a local CDW distortion,
which can in turn display similar dissipation phenomena to the long-range ordered CDW.

To model and understand in detail the effect of the tip on the CDW, and ultimately the source of the
dissipation peaks, we describe the density deformation as a simple elastic 
system interacting with an extended perturbation. Following well established literature 
\cite{Fukuyama1978, Lee1979}, we describe the CDW through phase and amplitude
order parameters, the latter assumed constant for small perturbations. The overall 
density (ions and electrons) takes the 
form $\rho(x)=\rho_0\cos(Qx+\phi(x))$, where $Q$ is the wavenumber and $\rho_0$ 
the constant amplitude. The energy as a function of the order parameter $\phi(x)$ 
for an external perturbing tip potential $V(x)$ then reads:
\begin{equation}\label{energy}
E[\phi(x)] = \int [(\nabla \phi(x))^2+V(x)\rho(x)]\mathrm{d}x.
\end{equation}
To keep it simple, this description assumes a unidirectional CDW, but the same mechanism 
should hold in our tridirectional NbSe$_2$ experimental example.
For a given external potential, the energy can be minimized to find the preferred phase shape 
along x. The main difference with previous treatments that were given for impurities \cite{Tucker1989}
is that in this case, due to the large size of the pendulum tip and swing, the potential $V(x)$ 
may extend over lengths 
longer than the characteristic CDW lengthscale $2\pi/Q  \sim$ 2.5 nm: this 
leads to new interesting effects. Additional care needs to be taken in choosing the dimensionality 
of the system: while the relevant tip-related behavior of the CDW is along x and restricted to
the surface, a full treatment will have to take into account the other dimensions, namely y in the
surface plane, and z into the underlying bulk, in order to obtain a realistic shape for the 
order parameter. While these aspects will be discussed in a separate work, here we will describe
the effect in a simple one dimensional model, which retains all the necessary features without
any of the complications connected with a higher dimensionality.

\begin{figure}[htb]
\centering
\includegraphics[width=0.48\textwidth]{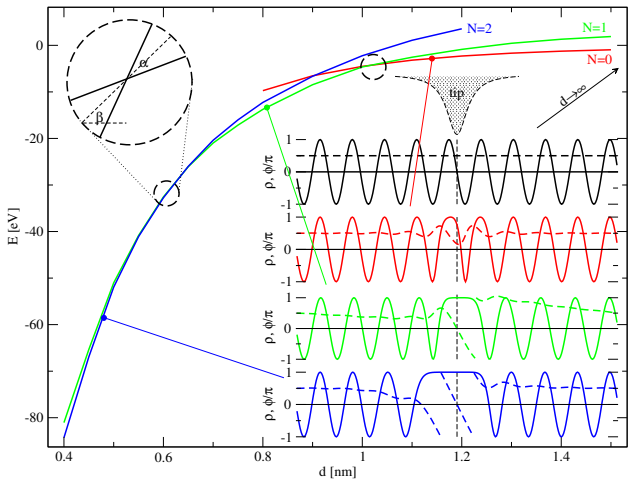}
    \caption{ Energy as a function of distance $d$ for the elastic CDW model under the tip perturbation.
Solutions are reported for different winding number $N$.
The dashed circles represent the crossover points.
\textbf{Inset:} Charge density $\rho$ (full lines) and phase $\phi$ (dashed lines) (see Eq.~\ref{energy}) along the
CDW direction for (from top to bottom) no perturbation, and increasing perturbations
with winding number $N=$ 0, 1, and 2 (at the marked distances $d$). The charge oscillation wavelength is $2\pi/Q$ 
and the position of the tip is indicated by the dashed vertical line. The $\phi$ boundary conditions are fixed at $\pi/2$.
}
   \label{fig3}
\end{figure}

An important role in understanding the dissipation mechanism is played
by the boundary conditions: while we will impose the phase to be fixed at the
boundaries (consistently with an overall pinned CDW as known experimentally), this value is only defined 
modulo $2\pi$. This leads to the existence of qualitatively different solutions
of the energy minimization problem, where (along the main CDW direction) the perturbation
may lead to a phase change $2\pi N$, with $N$ integer, between the pinned boundaries.
We will call $N$ the ``winding number'' of a given order parameter configuration.
We minimize the energy in the subspace of one given winding number and
observe its behavior as a function of the perturbation amplitude, controlled by the tip position above the substrate. 
We show this energy in Fig.~\ref{fig3} for a perturbation of Lorentzian form $V(x;d)$,
mimicking an approaching conical tip, as a function of the tip-sample distance $d$.

As energy curves for different winding numbers sharply cross at specific distances, marked
by dashed circles, we expect the CDW phase winding number to switch from one $N$ to the next $N \pm 1$ type of
deformation at these points. Moreover, as shown in the inset (phase and density along
the main CDW direction for different $N$), these solutions present a qualitative 
difference, akin to a symmetry difference, making a smooth transition between the two 
impossible. Passing through a crossover point will thus be accompanied by
a hysteresis cycle in the tip mechanics, which explains the giant size of dissipation peaks
despite the extremely low tip oscillation frequency as compared to dielectric relaxation 
frequency ($\sim$ MHz) measured in ac-conductivity experiments \cite{Tucker1989}.
To corroborate this picture, let us consider the height of the 
dissipation peaks it would produce, proportional to the area of the
hysteresis cycle. We expect this area to be related to the angle between
the energy curves $\alpha$ and the average slope at the crossing point $\beta$
(see the blow up in Fig.~\ref{fig3}). Considering that the area of the hysteresis cycle  while oscillating around this
point will behave like $W\propto\sin\alpha\cos^{-2}\beta$, we see from the shape of
the $E(d)$ curves that the variation of these two parameters have opposite 
effects on the height of the peak as the tip approaches the surface, since a 
decreasing $\alpha$ would decrease the hysteresis area, while an increasing
$\beta$ would increase it. This provides a possible explanation of the non-monotonic  
peak intensity behavior observed experimentally. A full treatment of the CDW related dissipation
materials should be followed by a  more thorough theoretical treatment, which will be reported elsewhere.
We can conclude here that the CDW phase slip model reproduces, at least qualitatively, the basic characteristics 
of the observed dissipation. 

It is important before closing to make connection
with earlier work, both experimental \cite{Saitoh2010},
where a single dissipation peak was reported on NbSe$_2$ and SrTiO$_3$,  and theoretical \cite{She2012},
where a collective resonance of spin-like centers was 
ingeniously invoked to explain it. Our findings of 
a multiplicity of interaction force-controlled peaks in NbSe$_2$, 
peaks systematically absent over NbS$_2$, suggests a different scenario.
The train of peaks is perfectly reproducible and ubiquitous over well ordered surfaces and
clean NbSe$_2$ bulks, making the existence and nature of these centers doubtful (see Fig.~4 of the Supplementary Material). 
The temperature evolution of the peaks suggests instead a strong connection with the CDW. 
A correspondence between each peak and the local pumping of a CDW  phase slip is 
made especially compelling by theoretically finding a first order energy crossing of states
of the CDW with different winding numbers, implying the possibility of a mechanical hysteresis 
cycle even at the very low tip vibration frequency, where an alternative viscous dissipation could be barely
detectable at best. 
In summary we believe we identified a promising mechanism -- coupling of a slider to a collective phase -- that 
is novel and deserves further experimental and theoretical study, extending to other systems with incommensurate
bulk phases. It should apply in particular to genuine charge-density-wave and spin-density-wave materials,
where the phase pumping should entail interesting surface flows of charge and spin under vibrating tips.

\section*{Methods}

\subsection*{Sample}

We used a high quality 2H-NbSe$_2$ single-crystal sample of size $5\times5\times0.8$
mm$^3$ with a $T_{\rm CDW}=33$~K determined from temperature dependent resistance
measurements which is in agreement with previous results \cite{Lin2005}. The
crystal was produced by means of standard chemical vapor transport technique
with iodine as the transport agent \cite{Kershaw1967}. The residuals
were washed off the surface with solvents. Sample was cleaved under UHV
conditions. 

\subsection*{Pendulum AFM microscope}

The experimental setup used for measuring friction at low temperature is
described in details in another publication \cite{gysin2011}. Here we note that
the probe
consisted of a soft cantilever (ATEC-CONT from Nanosensors) with spring constant
$k=120$~mN/m. The probe was suspended perpendicularly to the surface with an
accuracy of $1^\circ$ and operated in the so-called pendulum geometry in which
the tip vibrational motion is parallel to the sample surface. The oscillation
amplitude $A=5$~nm of the tip was kept constant using a \textit{Nanonis} phase
locked loop (PLL) feedback system. The cantilever tip is asymmetric as 
shown in Supplementary material. Due to this the perpendicular component of
oscillation amplitude $A_{\rm perp}=180$~pm exists as well. The cantilever was
annealed in UHV up to $700^\circ$~C for $12$ hours, which results in the removal of
water layers and other contaminants from both cantilever and tip. This is
confirmed by the improvement of the quality factor of the probe by almost 2
orders of magnitude \cite{nt_rast06}. It is also known that this long-term
annealing leads to negligible amounts of localized charges on the probing tip.
After annealing the cantilever was characterized by a resonance frequency $f=12$~kHz 
and a quality factor $Q=9\cdot 10^5$. This lead to the internal friction
$\Gamma_0 = \frac{k}{2 \pi f_0 Q}=1.7 \cdot 10^{-12}$~kg/s and the corresponding
dissipated power $P_0=\frac{\pi k A^2}{eQ}=65~\mu$W (at $6$~K). The sample and
cantilever temperatures were controlled by means of two different cryogenic
controllers (Model 340 and Model DRC-91C from Lakeshore Cryotronics, Inc., Westerville, Ohio,
USA).

\subsection*{Tuning-fork based STM/AFM}

The STM/AFM experiments were performed with a commercial qPlus STM/AFM
microscope (Omicron Nanotechnology GmbH) running at low temperature ($5$~K) under
UHV and operated by a Nanonis Control System from SPECS GmbH. We used a
commercial tuning fork sensor in the qPlus configuration (typical parameters:
resonance frequency $f_0\approx 25$~kHz, spring constant $k_0\approx1800$~N/m, 
typical quality factor $Q=35000$ at $5$~K). The oscillation amplitude was
always set to $200$~pm. The oscillation frequency $f(z)$ and excitation amplitude
$A_{\rm exc}$ were recorded simultaneously for different $V$ values applied to the
tungsten tip. The force extraction was done by means of the Sader and Jarvis
formula \cite{sader2004}. Additional $f(z)$ curve was taken (typical retraction
distance $\approx20$~nm) after the complete spectroscopic measurement in order
to estimate the long-rage vdW background. All STM images were recorded in the
constant current mode. 

\subsection*{Theoretical treatment}
The energy $E[\phi(x)]$ given in equation (\ref{energy}) was minimized numerically
through a simple conjugated gradient algorithm \cite{numrec} on a two-dimensional
discrete grid. The CDW was taken as unidirectional, while the other direction is
required for a realistic minimization of the phase. 
Different boundary conditions were enforced to obtain solutions with different 
winding number $N$, while within the grid the gradient was considered modulo 
$2\pi$ to allow for phase jumps when $N\neq0$. 
The external potential was taken as a Lorentzian depending on the tip-sample
distance $d$ as $V(r;d)=V_0(d)/(r^2+\sigma(d)^2)$, 
with $V_0(d)\propto d^{-1}$ and $\sigma(d)\propto d^2$. This shape is found to
be appropriate to model a conical tip interacting with the surface through van der 
Waals force. The parameters are estimated from a fit of the experimental overall
tip-sample interaction within the same model.

\subsection*{Acknowledgements}
FP, GES, and ET acknowledge research support by MIUR, through PRIN-2010LLKJBX\_001, 
by SNSF, through SINERGIA Project CRSII2 136287/1, and by the EU-Japan Project LEMSUPER. 
ET acknowledges ERC Advanced Research Grant
N. 320796  MODPHYSFRICT.
RB acknowledges financial support by the CNR program Short Term Mobility STM 2011.
ML, MK, RP, AB, and EM acknowledge financial support from the Swiss National Science Foundation (NSF), 
the SINERGIA Project CRSII2 136287/1 and the Swiss National Center of 
Competence in Research on “Nanoscale Science” (NCCR-NANO).GB acknowledges financial support from EPSRC, UK(EP/I007210/1).

\subsection*{Author contributions}
The samples was fabricated by GB. The idea was born out of discussion between ET, EM, RB, AG, RP and MK. 
The experiment was carried out by ML, MK and RP. RB and AG participated in sample preparation and pendulum AFM experiment.
The theoretical model was developed by FP, GES and ET. ET, EM, AB, FP, GES, RB, AG, ML, MK and RP were involved in interpretation, 
discussion and paper writing.

\pagebreak
\widetext
\begin{center}
\textbf{\large Supplementary material to CDW slips and giant frictional dissipation peaks at the NbSe$_2$ surface}
\end{center}
\setcounter{equation}{0}
\setcounter{figure}{0}
\setcounter{table}{0}
\setcounter{page}{1}
\makeatletter
\renewcommand{\theequation}{S\arabic{equation}}
\renewcommand{\thefigure}{S\arabic{figure}}
\renewcommand{\bibnumfmt}[1]{[S#1]}
\renewcommand{\citenumfont}[1]{S#1}

Additional 5 figures with captions:\\
- An asymmetric pendulum AFM tip.\\
- Dependence of resistivity as a function of temperature for NbSe$_2$ sample
used in the experiment.\\
- Temperature dependence of non-contact friction coefficient on NbSe$_2$
sample.\\
- Tip-sample position dependence of non-contact friction coefficient as the tip crosses the step edge
of NbSe$_2$ surface.\\ 
- Non-contact friction on NbS$_2$ - an intercalated transition metal compound
with no CDW signature.\\

\begin{figure}[hb!]
\centering
\includegraphics[scale=0.6]{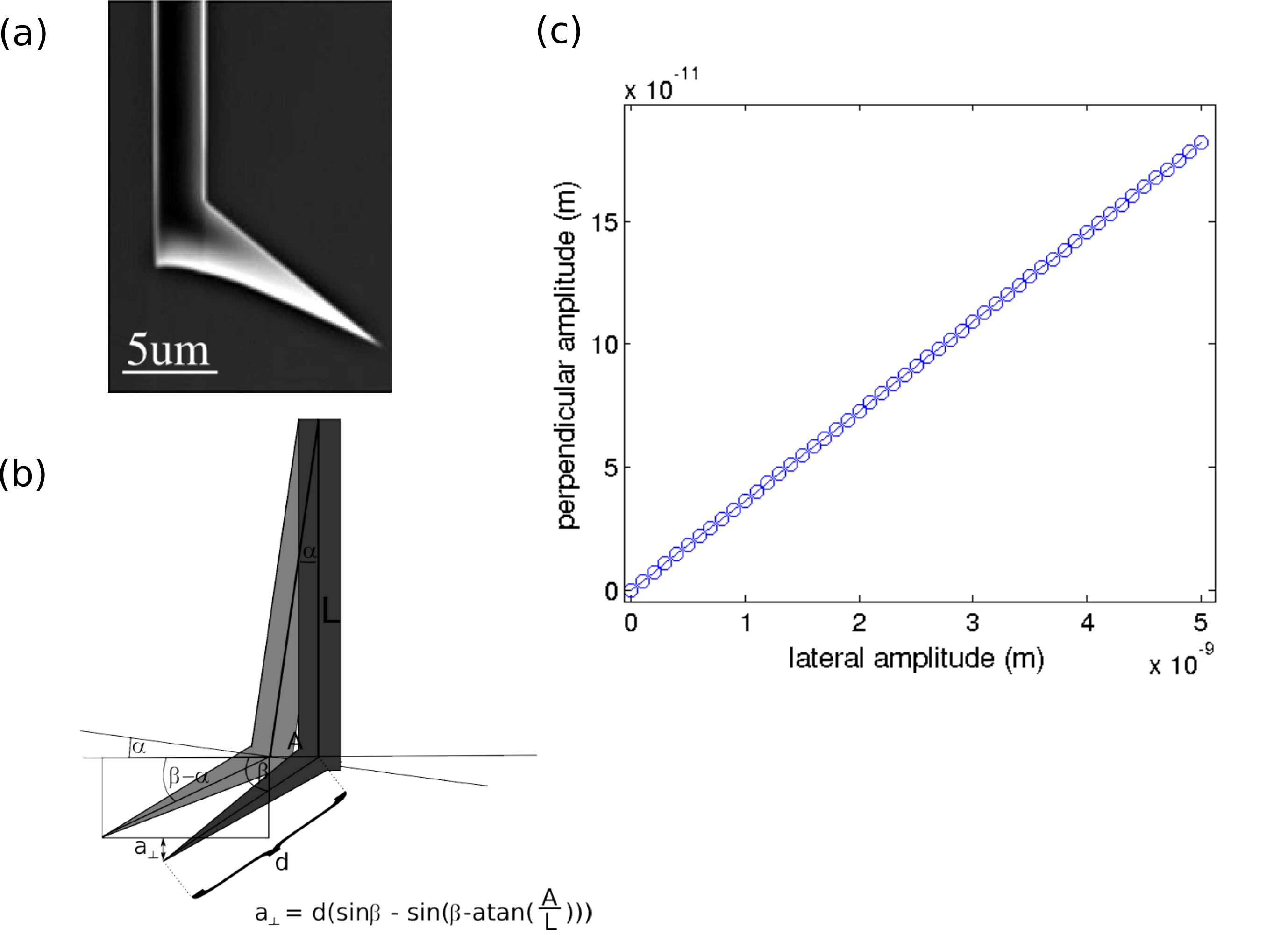}
   \caption{- \textbf{An asymmetric pendulum AFM tip.} \textbf{a}, SEM
micrograph of the Nanosensors - ATEC CONT cantilever tip. \textbf{b}, lateral
oscillation with amplitude $A$ leads to non-negligible perpendicular amplitude
of oscillation $a_\perp$ (L is the cantilever length). \textbf{c}, $a_\perp$
\textit{versus} $A$ for ATEC CONT cantilever.}
      \label{figS1}
\end{figure}

\begin{figure}[htb]
\centering
\includegraphics[scale=0.4]{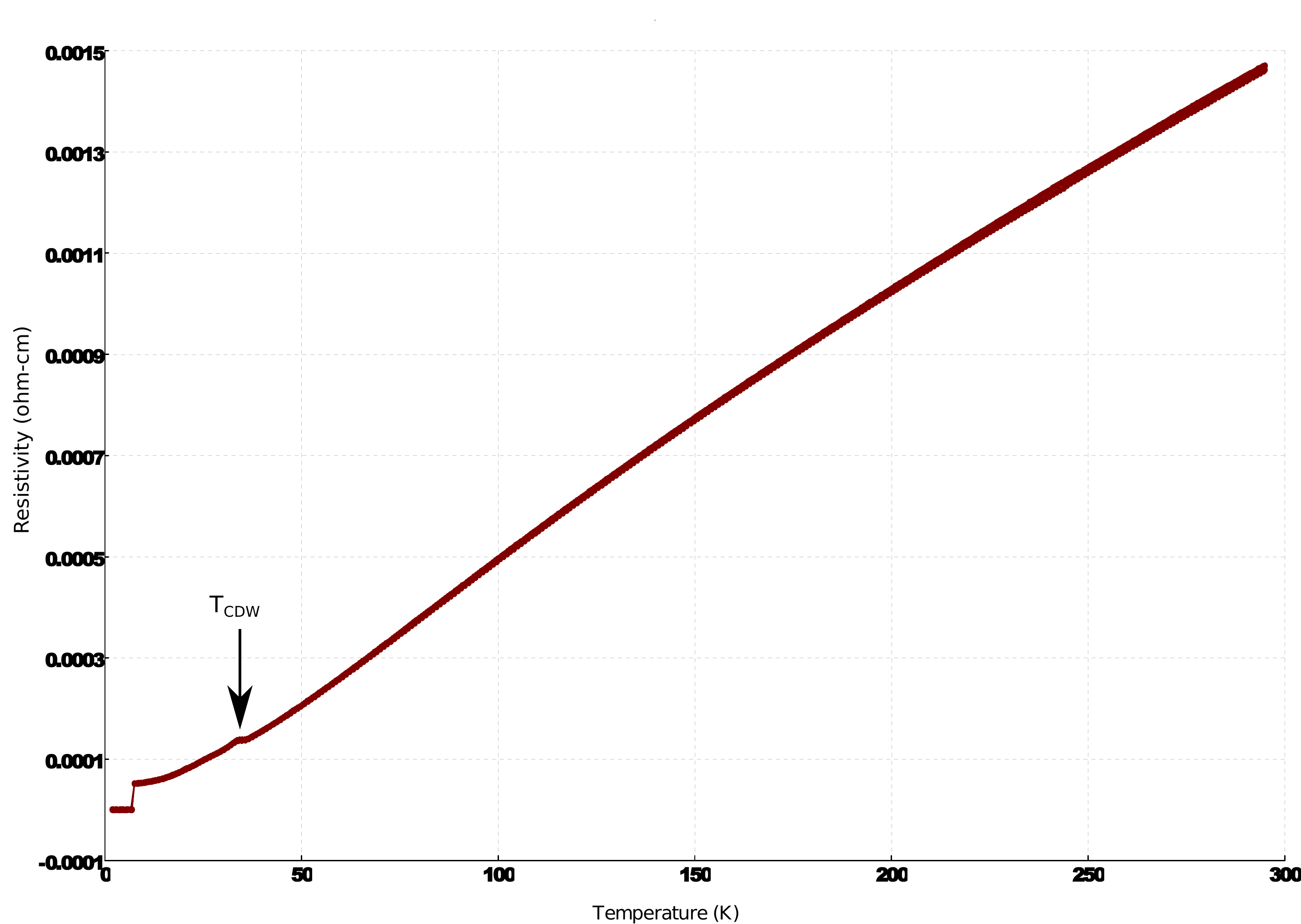}
   \caption{\textbf{Dependence of resistivity as a function of temperature for
NbSe$_2$ crystal used in the experiment.} At $T\sim33$~K sample undergoes a bulk
electronic CDW phase transition marked by an arrow. Transition to
superconducting state is also visible below $T=7.2$~K.}
   
   \label{figS2}
\end{figure}

\begin{figure}[htb]
\centering
\includegraphics[scale=0.7]{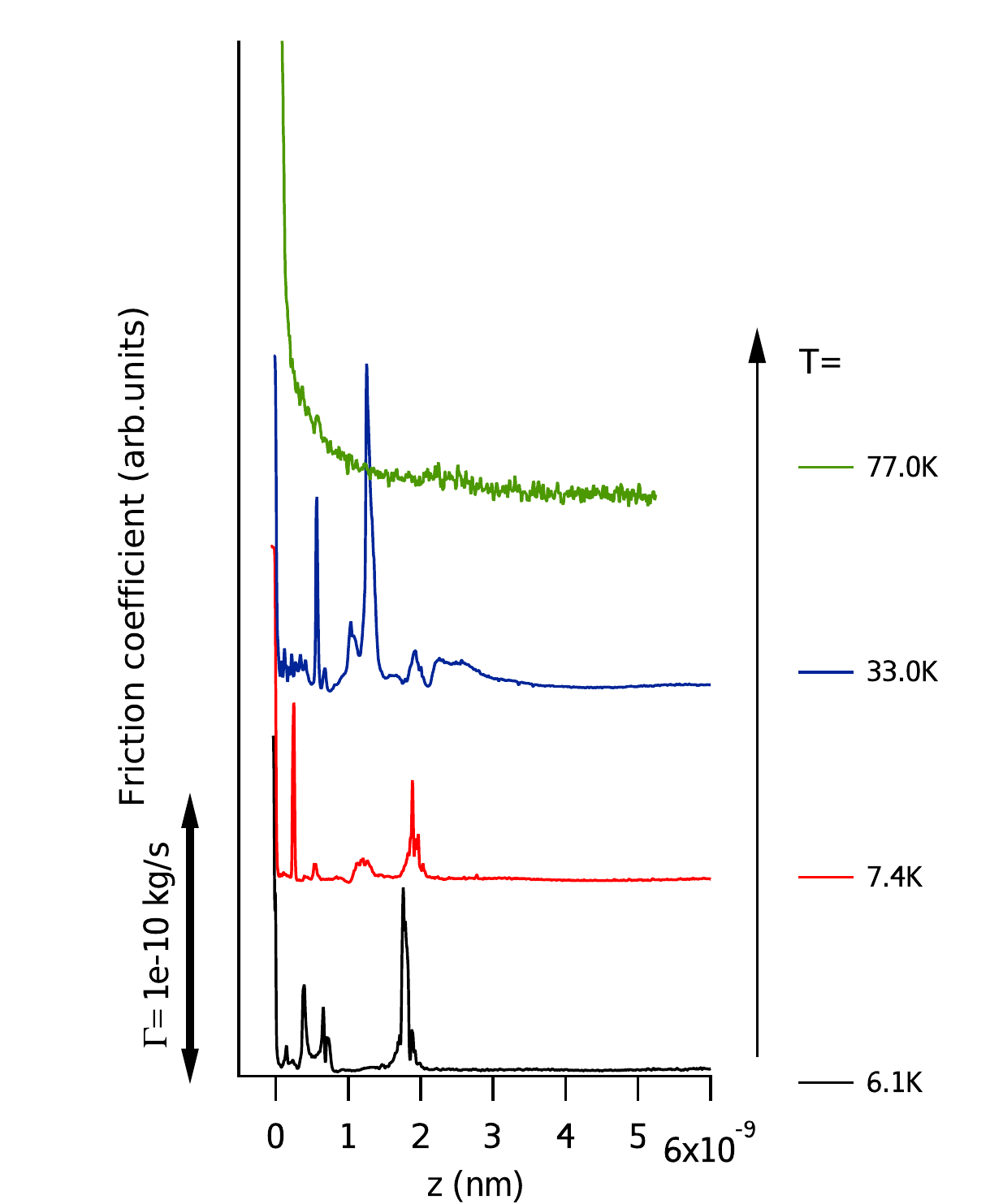}
   \caption{\textbf{Temperature dependence of non-contact friction coefficient
on NbSe$_2$ sample.} The huge dissipation maxima are observed up to $T=71 \pm
6$~K. Please note that because of technical reason it was impossible to perform T-dependent measurements at the same sample place. 
Mainly this effect is responsible for the strong variation of z-position of dissipation maxima. The effect of tip-sample position on dissipation spectra is shown in Fig.S4.}
   \label{figS3}
\end{figure}

\begin{figure}[htb]
\centering
\includegraphics[scale=0.4]{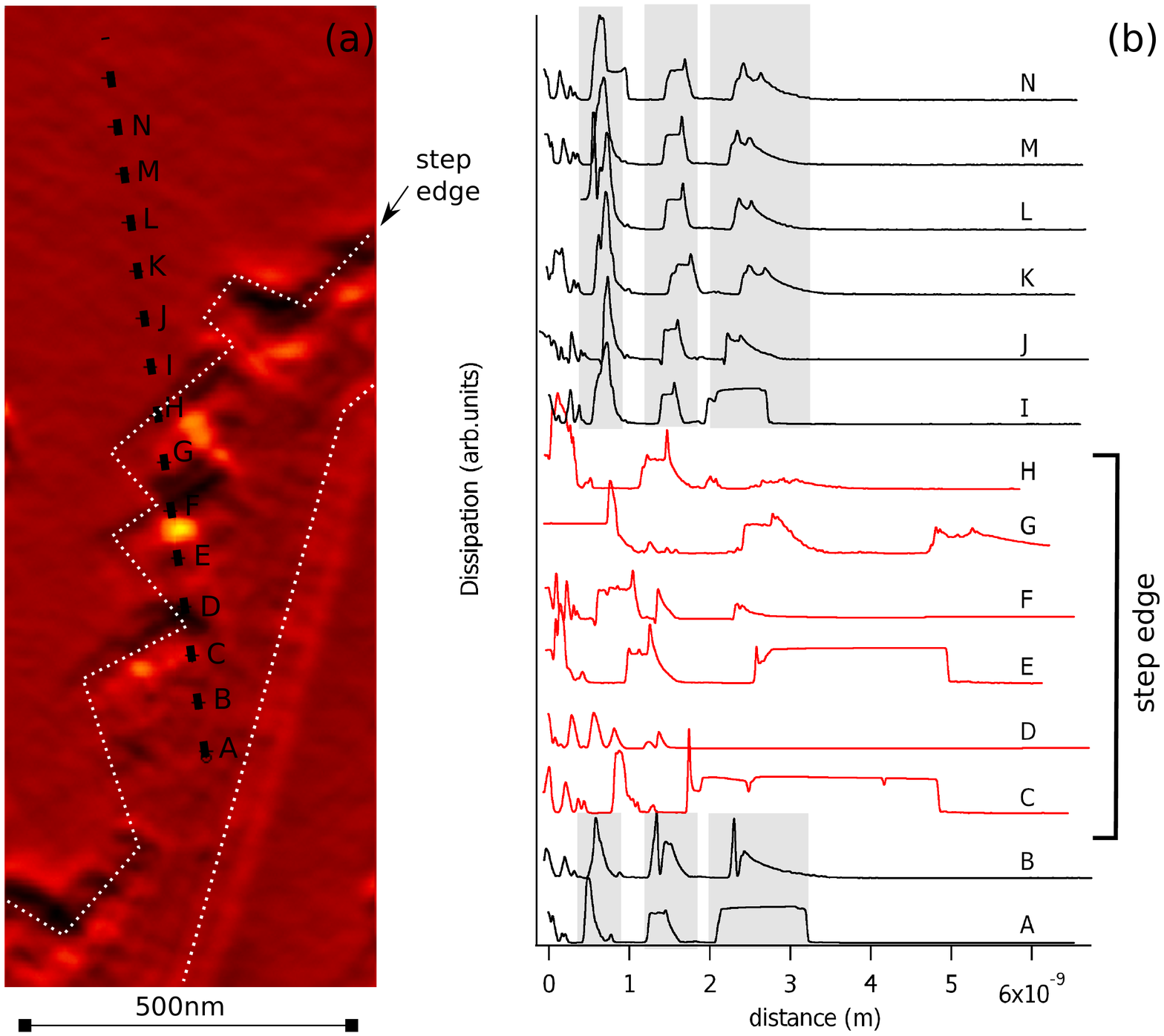}
   \caption{\textbf{Tip-sample position dependence of non-contact friction coefficient as the tip crosses the step edge
of NbSe$_2$ surface.} The tip moves from point A to N is it shown in Fig. (a). Corresponding huge dissipation maxima (shown in (b)) are observed over all the sample surface and the z-positions of the maxima scatters strongly when the tip is positioned close to the edge. Far from the step edge, on flat, clean place all the spectra (here highlighted with gray) are almost identical over $\mu m$ size surface area. Measurement was taken at T=6K.}
   \label{figS4}
\end{figure}

\begin{figure}[htb]
\centering
\includegraphics[scale=0.6]{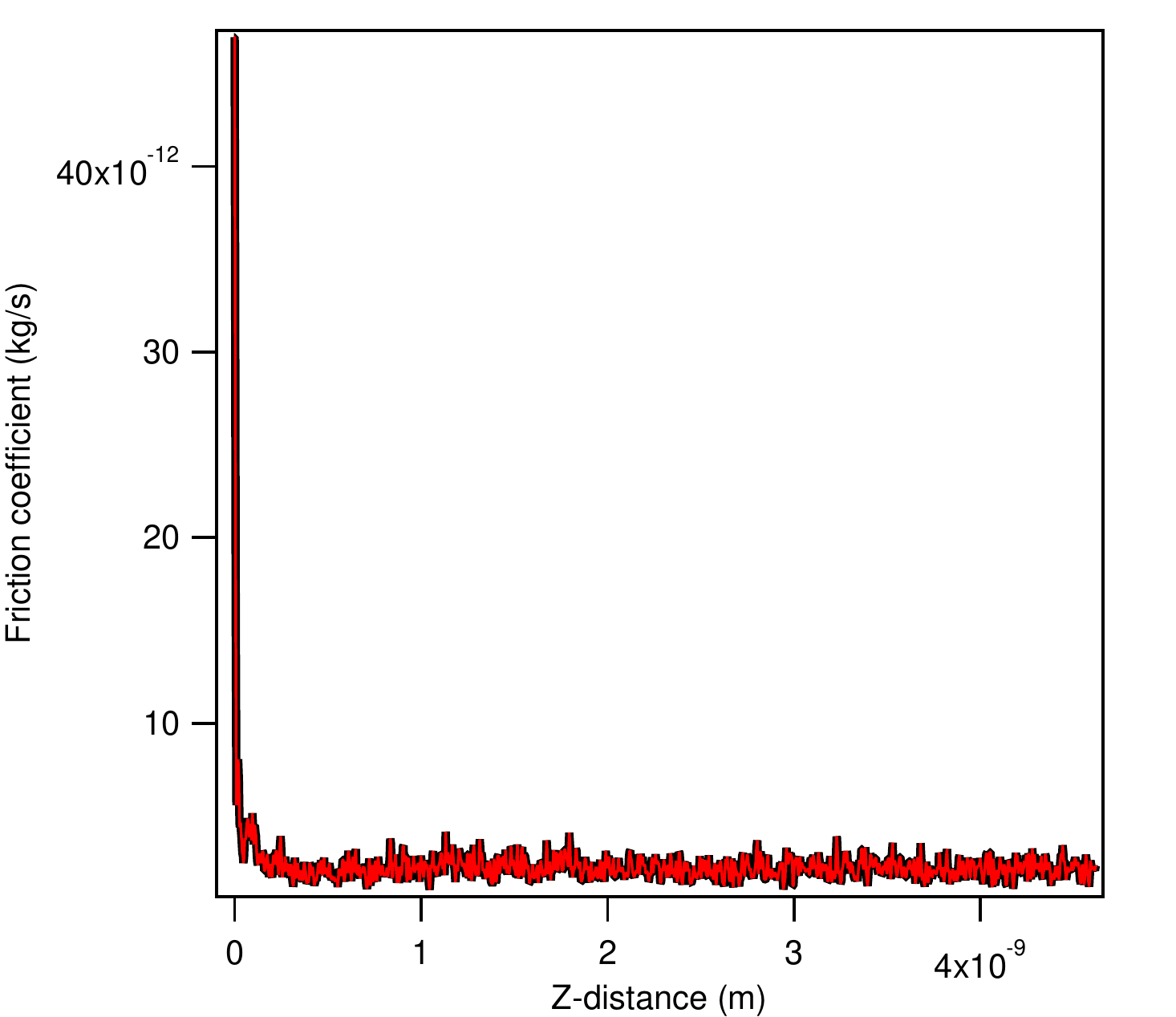}
   \caption{\textbf{Non-contact friction on NbS$_2$ - an intercalated transition
metal compound with no CDW signature.} Contrary to NbSe$_2$, the graph shows smooth increase of dissipation as the pendulum AFM tip
approaches the surface. Temperature of the measurement is $T=6$~K. 
The sample 2H-NbS$_2$ of size $0.84\times1.13\times0.98$ mm$^3$ has a superconducting transition at temperature $T\sim6K$ 
determined from temperature dependent conductance measurements. The sample was produced by means of standard chemical vapor transport 
technique with iodine as a transport agent. (``Stoichiometry, structure, and physical properties of niobium disulfide'' 
G. Fisher , M. J. Sienko Inorg. Chem., 19 (1), pp 39–43 (1980)). 
The sample was cleaved under UHV.}
   
   \label{figS5}
\end{figure}


\begin{thebibliography}{1}


\bibitem{Urbakh2010}
M. Urbakh and E. Meyer, \textit{Nanotribology: The renaissance of friction}.
Nature Mat. \textbf{9}, 8-10, (2010).

\bibitem{Lantz2009}
M.A. Lantz, D. Wiesmann, and B. Gotsmann, \textit{Dynamic superlubricity and the
elimination of wear on the nanoscale}, Nat. Nanotech. \textbf{4}, 586-591
(2009).

\bibitem{Socoliuc2006}
A. Socoliuc, E. Gnecco, S. Maier, O. Pfeiffer, A. Baratoff, R. Bennewitz, and E.
Meyer, \textit{Atomic-Scale Control of Friction by Actuation of Nanometer-Sized
Contacts}, Science \textbf{313}, 207-210 (2006).

\bibitem{Vanossi2013}
A. Vanossi, N. Manini, M. Urbakh, S. Zapperi, and E. Tosatti,
\textit{Modeling friction: from nanoscale to mesoscale},
Rev. Mod. Phys. \textbf{85}, 529 (2013).

\bibitem{Kisiel2011}
M. Kisiel, E. Gnecco, U. Gysin, L. Marot, S. Rast and E. Meyer,
\textit{Suppression of electronic friction on Nb films in the superconducting
state}, Nature Mat.
\textbf{10}, 119-122, (2011).

\bibitem{Volo2007}
A. I., Volokitin, B. N. J. Persson, H. Ueba,  \textit{Giant enhancement of
noncontact friction between closely spaced bodies by dielectric films and
two-dimensional systems}, Journ. Exp. Theor. Phys., \textbf{104}, 96-110,
(2007).

\bibitem{Stipe2001}
B.C. Stipe,  H.J. Mamin,  T.D. Stowe,  T.W. Kenny, and D. Rugar,
\textit{Noncontact Friction and Force Fluctuations between Closely Spaced
Bodies}, Phys. Rev. Lett. \textbf{87}, 096801 (2001).

\bibitem{Kuehn2006}
S. Kuehn, R.F. Loring, and J.A. Marohn, \textit{Dielectric Fluctuations and the
Origins of Noncontact Friction}, Phys. Rev. Lett. \textbf{96}, 156103 (2006).

\bibitem{Wilson1975}
J.A. Wilson, F.J. Di Salvo, and S. Mahajan, 
\textit{Charge-Density Waves and Superlattices in the Metallic Layered Transition 
Metal Dichalcogenides}, Adv. Phys. \textbf{24}, 117-201 (1975).

\bibitem{Moncton1975}  D. E. Moncton, J. D. Axe, and F. J. DiSalvo, \textit{Study 
of Superlattice Formation in $2H$-Nb${\mathrm{Se}}_{2}$ and $2H$-Ta${\mathrm{Se}}_{2}$ 
by Neutron Scattering}, Phys. Rev. Lett. \textbf{34}, 734 (1975).

\bibitem{Inosov2008}
D S Inosov and V B Zabolotnyy and D V Evtushinsky and A A Kordyuk and B Büchner
and R Follath and H Berger and S V Borisenko, \textit{Fermi surface nesting in
several transition metal dichalcogenides}, New Journal of Physics, \textbf{10},
125027, (2008).

\bibitem{Shen2008}
Shen, D. W. and Zhang, Y. and Yang, L. X. and Wei, J. and Ou, H. W. and Dong, J.
K. and Xie, B. P. and He, C. and Zhao, J. F. and Zhou, B. and Arita, M. and
Shimada, K. and Namatame, H. and Taniguchi, M. and Shi, J. and Feng, D. L.,
\textit{Primary Role of the Barely Occupied States in the Charge Density Wave
Formation of ${\mathrm{NbSe}}_{2}$}, Phys. Rev. Lett. \textbf{101}, 226406,
(2008).

\bibitem{Borisenko2009}
Borisenko, S. V. and Kordyuk, A. A. and Zabolotnyy, V. B. and Inosov, D. S. and
Evtushinsky, D. and B\"uchner, B. and Yaresko, A. N. and Varykhalov, A. and
Follath, R. and Eberhardt, W. and Patthey, L. and Berger, H., \textit{Two Energy
Gaps and Fermi-Surface ``Arcs'' in ${\mathrm{NbSe}}_{2}$}, Phys. Rev. Lett.
\textbf{102}, 166402, (2009).
  
\bibitem{Weber2011}
Weber, F. and Rosenkranz, S. and Castellan, J. P. and Osborn, R. and Hott, R.
and Heid, R. and Bohnen, K. P. and Egami, T. and Said, A. H. and Reznik, D.,
\textit{Extended Phonon Collapse and the Origin of the Charge-Density Wave in
$2H-{\mathrm{NbSe}}_{2}$}, Phys. Rev. Lett. \textbf{107}, 107403, (2011).

\bibitem{Pasupathy2013}
S.P. Chockalingam, C.J. Arguello, E.P. Rosenthal, L. Zhao, C. Gutiérrez, 
J.H. Kang, W.C. Chung, R.M. Fernandes, S. Jia, A.J. Millis, R.J. Cava, 
A.N. Pasupathy, \textit{Visualizing the Charge Density Wave Transition in 
2H-NbSe2 in Real Space}, arXiv:1307.2282 [cond-mat.str-el], (2013).

\bibitem{Guillamon2008}
Guillam\'on, I. and Suderow, H. and Vieira, S. and Cario, L. and Diener, P. and
Rodi\`ere, P., \textit{Superconducting Density of States and Vortex Cores of
2H-${\mathrm{NbS}}_{2}$}, Phys. Rev. Lett. \textbf{101}, 166407, (2008).

\bibitem{Soumyan2013}
A. Soumyanarayanan, M.M. Yee, Y. He, J. van Wezel, D.J. Rahn, K. Rossnagel,
E.W. Hudson, M.R. Norman, and J.E. Hoffman, \textit{Quantum phase transition from 
triangular to stripe charge order in NbSe2}, PNAS \textbf{110}, 1623, (2013).

\bibitem{Cleveland1998}
J. P. Cleveland, B. Anczykowski, A. E. Schmid, and V. B. Elings,
\textit{Energy dissipation in tapping-mode atomic force microscopy}
Appl. Phys. Lett. \textbf{72}, 2613 (1998).


\bibitem{sader2004}
John E. Sader and Suzanne P. Jarvis, \textit{Accurate formulas for interaction
force and energy in frequency modulation force spectroscopy}, Applied Physics
Letters \textbf{84}, 1801-1803, (2004).


\bibitem{Saitoh2010}
K. Saitoh, K. Hayashi, Y. Shibayama, K. Shirahama, \textit{Gigantic Maximum of
Nanoscale Noncontact Friction}, Phys. Rev. Lett.\textbf{105}, 236103, (2010).


\bibitem{Fukuyama1978}
Fukuyama, H. and Lee, P. A., \textit{Dynamics of the charge-density wave. I.
Impurity pinning in a single chain}, Phys. Rev. B \textbf{17}, 535-541, (1978).


\bibitem{Lee1979}
Lee, P. A. and Rice, T. M., \textit{Electric field depinning of charge density
waves}, Phys. Rev. B \textbf{19}, 3970-3980, (1979).

\bibitem{Tucker1989}
Tucker, J. R. \textit{Impurity pinning of sliding charge-density waves}, Phys.
Rev. B\textbf{40}, 5447-5459, (1989).


\bibitem{She2012}
She, Jian-Huang and Balatsky, Alexander V., \textit{Noncontact Friction and
Relaxational Dynamics of Surface Defects}, Phys. Rev. Lett.\textbf{108}, 136101,
(2012).


\bibitem{Lin2005}
Lin-jun Li and Zhu-an Xu and Jing-qin Shen and Li-min Qiu and Zhi-hua Gan,
\textit{The effect of a charge-density wave transition on the transport
properties of 2H-NbSe2}, Journal of Physics: Condensed Matter \textbf{17}, 493,
(2005).

\bibitem{Kershaw1967}
Kershaw, R. and Vlasse, Marcus and Wold, Aaron, \textit{The preparation of and
electrical properties of niobium selenide and tungsten selenide}, Inorganic
Chemistry \textbf{6}, 1599-1602, (1967).

\bibitem{gysin2011}
U. Gysin and S. Rast and M. Kisiel and C. Werle and E. Meyer, \textit{Low
temperature ultrahigh vacuum noncontact atomic force microscope in the pendulum
geometry}, Review of Scientific Instruments \textbf{82}, 023705, (2011).

\bibitem{nt_rast06}
S. Rast, U. Gysin, P.Ruff, C. Gerber, E. Meyer and D.W. Lee, \textit{Force
microscopy experiments with ultrasensitive cantilevers}, Nanotech. \textbf{17},
189-194 (2006).

\bibitem{numrec}
W.H. Press, S.A. Teukolsky, W.T. Vetterling, and B.P. Flannery,
\textit{Numerical Recipes: The Art of Scientific Computing (3rd ed.)},
Cambridge University Press (2007).

\end{thebibliography}
\end{document}